\documentclass[12pt]{article}
\usepackage{cite}
\begin{document}
\title{Thermodynamic Gravity and the Schr\"odinger Equation}
\author{{\bf Merab Gogberashvili}\\
Andronikashvili Institute of Physics\\
6 Tamarashvili Street, Tbilisi 0177, Georgia\\
and\\
Javakhishvili State University\\
3 Chavchavadze Avenue, Tbilisi 0128, Georgia\\
\\
{\sl E-mail: gogber@gmail.com}}
\maketitle
\begin{abstract}
We adopt a formulation of the Mach principle that the rest mass of a particle is a measure of it's long-range collective interactions with all other particles inside the horizon. As a consequence, all particles in the universe form a 'gravitationally entangled' statistical ensemble and one can apply the approach of classical statistical mechanics to it. It is shown that both the Schr\"odinger equation and the Planck constant can be derived within this Machian model of the universe. The appearance of probabilities, complex wave functions, and quantization conditions is related to the discreetness and finiteness of the Machian ensemble.
\end{abstract}

\vskip 0.5cm


\section{Introduction}

There exist three types of field theoretical descriptions successfully working in their own areas: quantum mechanics, quantum field theory, and general relativity. However, the problems arise, when one tries to establish bridges between these descriptions. There are well known difficulties in foundations of relativistic quantum mechanics, the definition of quantum fields in curved space-times, or quantization of gravity. A new hope to establish a unifying approach to main field theoretical descriptions is provided by the approach appealing to the notions of thermodynamics. For instance, thermodynamic arguments have been used in the black hole physics \cite{BH}, discovery of Unruh temperature \cite{Unr}, establishment of the AdS/CFT correspondence \cite{AdS}, derivation of the Einstein \cite{Ein} and Maxwell \cite{Wang} equations, in the recent attempt to interpret the Newtonian gravity as an entropic force \cite{Ver}, and other discussions of the "emergent gravity" \cite{Pad,Hu}. Besides, the analogies with classical statistical mechanics and thermodynamics have been underlying some recent discussions of the foundations of quantum mechanics \cite{Wet,Gro}.

These surprising relationships are still regarded as just analogies, since the conception that classical field theory is a consequence of the laws of thermodynamics and statistical physics rises many questions. The main question is the origin of the microscopic degrees of freedom responsible for the thermodynamic quantities like temperature or entropy in these approaches. Besides, when considering physical theories as emergent phenomena from some thermodynamic processes, it seems necessary to modify the notion of locality for the underlining theory, since, at least in quantum mechanics, it is known that models with hidden variables lead to problems with locality \cite{NonLoc}. Another problem is how to explain the relativity principle, because in general, the setting of the thermodynamical equilibrium should result in the appearance of a global preferred frame (even without the Galilean symmetry).

Note also, that if one tries to give a thermodynamic interpretation to the quantum properties of nature, one should be able to provide an interpretation of the Planck constant $\hbar$. At the present a full understanding of the appearance of $\hbar$ in the formulas of Hawking and Unruh temperatures, or in the formula of a black hole energy, is missing.

Here we propose solutions of some problems of thermodynamical approaches within the Machian model introduced in our previous papers \cite{Gog1,Gog2,Gog3,Gog4,Gog5}, see also \cite{Win}. The feature of our model, which in our opinion should be added to the thermodynamical approach, is the assumption of non-locality and the existence of a global universal frame in the underlying theory. In fact, the search for an acceptable framework for quantum gravity already has motivated several authors to introduce non-local corrections for special \cite{nonSR} and general \cite{nonGR} relativity theories. The non-locality in gravity can be understood as a manifestation of the Mach principle, which in its utmost generality assumes that in any mechanical act the interaction with the entire universe is involved.

Several formulations of the Mach principle can be found in the literature \cite{Mach}. The usual formulation, which leads to the anisotropy of the rest mass of a particles due to the influence of nearby massive objects (like the Galaxy), has been ruled out by experiments \cite{exp}. In fact, the naive Machian proposition that the mass parameter can be altered by "distant stars" is based on the assumption that kinematics, or local space-time, is independent of the surrounding universe. However, the lesson of General Relativity has been that the description of space-time geometry, and hence the kinematics itself, depend on the distribution of matter. Therefore, the influence of the whole universe on the local physics, or Mach's principle, should not be described in terms of the mass parameters alone, but rather in terms of more fundamental quantities such as action or energy.

We adopt the thermodynamical-statistical formulation of the Mach principle \cite{Gog1,Gog2,Gog3,Gog4,Gog5} that the rest mass of a particle is a measure of its long-range collective gravitational interactions with all other particles inside the horizon. The assumption that the whole universe is involved in local interactions effectively weakens the observed strength of gravity by a factor related to the number of particles in the universe and it can explain the famous 'hierarchy problem' in particle physics and cosmology \cite{Gog1}. In the Machian model all the particles in the universe are 'gravitationally entangled' and form a statistical ensemble \cite{Gog3}, which singles up the fundamental cosmological frame \cite{Gog4}. In spite of this, in IR our model is compatible with other cosmological and gravitational theories. For instance, in \cite{Gog2,Gog5} it is argued that the Machian model imitates basic features of special and general relativity theories, such as the relativity principle, the weak equivalence principle, the local Lorentz invariance, and obeys the so-called signal locality, i.e. the matter cannot propagate faster than light \cite{Gog2}.

Non-locality is also known to be an essential part of quantum mechanics. It is hidden in usual interpretations, in the notion of collapse of the wave function, but it is explicit in formulation of quantum mechanics which appeal to the notions from classical physics (see, e.g. \cite{Nic}). It is natural to assume that the non-locality in the Machian model of gravity and that in quantum mechanics have common origin. In this paper we show that by using classical statistical mechanics for the Machian universe one can obtain both the Schr\"odinger equation and the main features of quantum mechanics, including an interpretation of $\hbar$.

There exist several other attempts to derive the Schr\"odinger equation from classical physics: using the non-equilibrium thermodynamics \cite{Gro}, within the hydrodynamical interpretation of quantum mechanics \cite{Mad}, in Nelson's stochastic mechanics \cite{Nel}, using the hidden-variable theory of de Broglie and Bohm \cite{Bohm}, or from the 'exact uncertainty principle' \cite{Ha-Re}. All those models usually use the polar form of the quantum mechanical wave function
\begin{equation} \label{Psi}
\Psi = |\Psi| e^{i {\cal S}/\hbar}~,
\end{equation}
where ${\cal S}$ is identified with the particle's action, or it's Hamilton's principle function. Then the Schr\"odinger equation is split into the system of nonlinear equations for the functions $|\Psi|$ and ${\cal S}$ and the attempts are made to derive those equations using the concepts of classical physics.

In spite of the intriguing analogies of the quantum Schr\"odinger equation with various equations admitting a classical interpretation, there are also some well-known differences. First, within the classical physics it is difficult to justify that the normalization constant $\hbar$ in (\ref{Psi}) coincides with the Planck constant. Another problem is that the classical physics is deterministic, while the usual form of quantum mechanics is fundamentally probabilistic. The attempts of classical interpretation of the components of wave function also face the problems of non-linearity and non-locality of the corresponding equations, whereas the advantages of the standard formulation of quantum mechanics using the complex wave function are related to the simplicity and linearity of the Schr\"odinger equation.

In this paper we demonstrate that within the Machian model of the universe \cite{Gog1,Gog2,Gog3,Gog4,Gog5} these problems can be resolved. For instance, the Planck constant can be understood as the action of the universe per particle of the Machian ensemble. Also, using the formalism of classical statistical mechanics, we derive the equations of motion of particles of this ensemble and show that they are equivalent to the quantum Schr\"odinger equation, and also clarify the origin of the quantization condition.


\section{Parameters of the Machian model and $\hbar$}

The specifics of our Machian model \cite{Gog1,Gog2,Gog3,Gog4,Gog5} is that each particle in the universe is interacting with the non-local potential resulting from the collective gravitational interaction of all $N$ particles in the universe. Consequently, the universe can be considered as a statistical ensemble of 'gravitationally entangled' particles. The non-local Machian interaction gives rise to what we usually perceive as classical space-time. The universal constant of the speed of light, $c$, originates in the non-local Machian potential of the whole universe, $\Phi$, acting on each particle of the world ensemble:
\begin{equation} \label{Phi}
c^2 = - \Phi = \frac{2M_U G}{R}~,
\end{equation}
where $M_U$ and $R$ are the total mass and the radius of the universe, respectively, and $G$ is the Newton constant (c.f. \cite{Sci}). This 'universal' potential $\Phi$, and thus $c$, can be regarded as constants, since, according to the cosmological principle, the universe is isotropic and homogeneous on the horizon scales $R$. Let us emphasize that (\ref{Phi}) is equivalent to the critical density condition in relativistic cosmology:
\begin{equation} \label{rho}
\rho_{c} = \frac{3M_U}{4\pi R^3} = \frac{3H^2}{8\pi G} ~,
\end{equation}
where $H \sim c/R$ is the Hubble constant.

Using the observed values of $c$, $G$ and $H$ in (\ref{rho}), or in (\ref{Phi}), we can estimate the total mass of the universe:
\begin{equation} \label{M}
M_U \sim \frac{c^3}{2 GH} \approx 10^{53}~ kg~.
\end{equation}

The relation (\ref{Phi}) allows us to formulate the Mach principle which relates the origin of inertia of a particle, or its rest energy, to the particle's interactions with the whole universe:
\begin{equation} \label{mPhi}
E = mc^2 = - m\Phi~,
\end{equation}
where $m$ is the mass parameter describing the particle's inertia, which in general is not constant.

The universal Machian potential in eq. (\ref{mPhi}) takes into account the contribution of the collective gravitational interactions between all $N$ particles inside the horizon. Namely, since each particle interacts with all other $(N-1)$ particles, and the mean separation in the interacting pairs is $R/2$, the total Machian energy consists of $N(N-1)/2$ terms of magnitude $\approx 2G m^2/R$. Then, for very large $N$, the Machian energy of a single particle which interacts with the total Machian potential of the universe $\Phi$ is given by:
\begin{equation} \label{E}
E \approx N^2 \frac {Gm^2}{R} ~.
\end{equation}
Correspondingly, the contribution of the collective Machian interactions to the total mass of the universe is:
\begin{equation} \label{M=N2m}
M_{Mach} \approx \frac{1}{2} N^2 m ~,
\end{equation}
so that the total mass of the universe $M_U\sim M_{Mach}$ is of the order of $N^2$ and not $\approx Nm$.

Because of the finite number of particles inside the horizon and the existence of the maximal speed $c$, any movement of the particles of the "gravitationally entangled" world ensemble results in a delayed response of the whole ensemble. The response time of the universe to the motion of a quantum particle is estimated as follows:
\begin{equation} \label{Deltat}
\Delta t \sim \frac{R}{Nc} \sim \frac{1}{NH}~.
\end{equation}
Note that $\Delta t$ is much shorter than e.g. the mean free motion time of particles with the mean separation $\sim R / N^{1/3}$ in a dilute gas, because, as a result of the non-local Machian interactions of all particles, the effective mean separation between particles in the world ensemble is much shorter: $\sim R/N$.

As a consequence of the delayed response of the universe, any mechanical process in the world ensemble will be accompanied by the exchange of at least the minimal amount of action $A = m c^2 \Delta t$, which we identify with the Planck's action quantum \cite{Gog3}:
\begin{equation} \label{A}
A = -\int dt~ E \approx - m c^2 \Delta t = 2\pi\hbar ~.
\end{equation}
If the particle before and after $\Delta t$ is 'free', or has uniform velocities, the universality of Machian energy in inertial frames (where the universe looks spherically symmetric and $N$, and hence, $E$ is conserved) leads to the principle of least action \cite{Gog2}:
\begin{equation} \label{deltaA}
\delta A = 0~.
\end{equation}

From (\ref{M}), (\ref{Deltat}) and (\ref{A}) we can estimate the total action of the universe:
\begin{equation} \label{A_U}
A_U = \frac{M_U c^2}{H} \approx \frac{N^3}{2} A ~,
\end{equation}
and the number of typical particles in it:
\begin{equation} \label{N}
N \approx \left(\frac{2 A_U}{A} \right)^{1/3}
\approx \left(\frac{M_U c^2}{\pi \hbar H} \right)^{1/3}
\approx 10^{40} ~.
\end{equation}
This number, one of the main parameters of the Machian model, is known to have appeared in a different context in Dirac's 'large numbers' hypothesis, which points to the existence of a deep connection between the micro and macro physics \cite{large}.

Using the value (\ref{N}) and eqs. (\ref{M=N2m}) and (\ref{M}), we can also estimate the mass of a typical particle in our simplified Machian universe:
\begin{equation} \label{m}
m \approx {\frac{2 M_U}{N^2}} \approx 2 \times 10^{-27}~ kg
\approx 1 ~ GeV~c^{-2}~,
\end{equation}
which appears to be of the order of magnitude of the proton mass. This estimation is also consistent with (\ref{Deltat}), as
\begin{equation} \label{Deltat=}
\Delta t \approx \frac{1}{NH}
\approx 0.5\times 10^{-22}~s \approx \hbar ~ GeV^{-1}~.
\end{equation}
Hence, a typical stable heavy particle, the proton, can be considered as a typical particle forming the gravitating world ensemble in our simplified one-component Machian universe. Or, vice versa, we could postulate that the typical mass of a particle forming the world ensemble equals to the proton mass (i.e. a typical stable baryon), and then use (\ref{Deltat}) in order to obtain the mean action per particle (\ref{A}), which then exactly coincides with the Planck constant.

The energy balance equations like (\ref{mPhi}) exhibit the exact conservation of energy in the Machian universe, and thus, they may help to avoid the problems with energy in Einstein's General Relativity \cite{Gog5}. Those energy balance conditions assume that the non-local Machian gravitational interaction with the universe is the source of all kinds of local energy of a particle, i.e. the total (Machian plus local) energy of any object vanishes, if the gravitational energy is considered as negative and all other forms of energy are assumed to be positive. For example, eq. (\ref{mPhi}) is equivalent to:
\begin{equation} \label{balance}
mc^2 + m\Phi = 0~.
\end{equation}
Consequently, the total energy of the whole universe also vanishes. This point of view appears to be preferable in cosmology \cite{Haw}, since in this case the universe can emerge without violating the energy conservation.

Let us also note that since the number of particles in the universe $N$ is integer, the typical feedback time (\ref{Deltat}) is discrete. Hence we can introduce a fundamental frequency of oscillations of a particle due to its interaction with the world ensemble:
\begin{equation}
\omega = \frac{1}{2\pi\Delta t}~,
\end{equation}
Then this quantity and $\hbar$ can be used in the definition of the energy of a particle,
\begin{equation} \label{homega}
E = \hbar \omega ~,
\end{equation}
which appears as an alternative to (\ref{mPhi}). The definition of the energy given by (\ref{mPhi}), with $\Phi = - c^2$, is convenient for massive objects. Whereas the alternative definition (\ref{homega}), using $\hbar$, is more useful for elementary particles, since in the zero-mass limit the description (\ref{mPhi}) fails.

The energy balance conditions such as (\ref{balance}) provide a framework for the treatment of the Machian influence of the universe on the local physics and the description of dynamical and kinematical parameters. For example, in \cite{Gog2} it was demonstrated that, in spite of the existence of a preferred frame, the approach based on the Machian energy balance equations is able to imitate basic features of the Special Relativity theory. Namely, the relativity principle emerges from the fact that in the homogeneous Machian universe there exists a class of privileged observers which have constant velocities with respect to the preferred frame, for which the universe looks spherically symmetric (see the standard definition of inertial frames e.g. in \cite{Wei}). However, in order to introduce the relativity principle into the model with a preferred frame, we have to pay with the velocity dependence of the inertia of particles. The Machian model of the universe is also capable of reproducing the Einstein equations and standard predictions of General Relativity \cite{Gog5}, as it is compatible with the weak equivalence principle, local Lorentz invariance, and the local position invariance, which are sufficient to describe gravity in terms of the geometry of space-time \cite{Wil}.


\section{Equations for the 'universal' ensemble}

Our aim now is to show how the Machian approach \cite{Gog1,Gog2,Gog3,Gog4,Gog5} can help to clarify the nature of space-time and solve the problems of classical interpretations of the Schr\"odinger equation \cite{Gro,Mad,Nel,Bohm,Ha-Re}.

First, because of the assumption of non-local Machian correlations extending up to the horizon scale all particles in the universe can be considered as members of a world statistical ensemble. This clarifies the origin of the microscopic degrees of freedom and allows us to describe the universe using the laws of statistical mechanics.

As in the previous models attempting a classical interpretation of the Schr\"odinger equation \cite{Gro,Mad,Nel,Bohm,Ha-Re}, let us describe our Machian ensemble by means of the distribution function
\begin{equation}\label{P}
P_n(t) = e^{2 K_n(t)/\hbar}~, ~~~~~~~~ \sum_n P_n (t) = 1 ~,
\end{equation}
and the Hamilton's principal function ${\cal S}_n(t)$, where the label $n$ runs over $1,2,...,N$ and $N$ is the total number of particles in the ensemble. For convenience, the logarithm of the distribution function, $K_n(t)$, is introduced in (\ref{P}).

The normalization parameter in (\ref{P}) is naturally identified with the Planck constant $\hbar$, since in our model it represents the portion of the action of the whole universe per the particle of the 'universal' ensemble. This interpretation of $\hbar$ overcomes another difficulty of the models of classical interpretations of quantum mechanics.

By means of $P_n(t)$ we can write down the action integral (\ref{A}) for a particle in the ensemble:
\begin{equation} \label{AP}
A = - \sum_n\int dt~P_n E ~.
\end{equation}
As the number of particles in the universe is huge, it is convenient to introduce the continuous limit by replacing the distribution function $P_n(t)$ with the probability density $P(t,x^i)$ ($i = 1,2,3$), where $x^i$ are continuous variables labeling the individual particles. Then the normalization condition (\ref{P}) takes the form:
\begin{equation}
\int d^3x ~P = \int d^3x ~e^{2 K(t,x^i)/\hbar} = 1 ~.
\end{equation}
Correspondingly, for the action functional (\ref{AP}) we have:
\begin{equation} \label{Ax}
A = - \int dtd^3x~P E ~.
\end{equation}
This means that while in UV we deal with the discrete ensemble and the single parameter of time, the description of the system in IR involves the continuous space coordinates $x^i$.

Note that the formula (\ref{Ax}) is written for a particle at rest with respect to the preferred frame, where the Hamilton's principle function is $ {\cal S}_0 \sim - Et$. In general, the action of a particle from the ensemble can be written in the form:
\begin{equation} \label{Ai}
A = \int dtd^3x~ P \left( \frac{\partial {\cal S}}{\partial t} + \frac{\nabla_i {\cal S} \nabla^i {\cal S}}{2m} + V \right) - \oint_L dl^i \nabla_i {\cal S}_v |^{t_2}_{t_1}~,
\end{equation}
where $V(t,x^i)$ is some local potential. The 3-momentum of the particle $p^i$ is related with the Hamilton's principle function ${\cal S}$ by the standard classical formula:
\begin{equation} \label{p}
p^i = \nabla^i {\cal S}~.
\end{equation}
If one imposes the requirement that at the times $t_2$ and $t_1$ the configuration of the system is prescribed, the last term in the action integral (\ref{Ai}) drops out and its variation with respect to $P$ and ${\cal S}$ gives rise to the Hamilton-Jacobi equation
\begin{equation} \label{HJE}
\frac{\partial {\cal S}}{\partial t} + \frac{\nabla_i {\cal S} \nabla^i {\cal S}}{2m} + V = 0
\end{equation}
and the continuity equation
\begin{equation} \label{cont}
\frac{\partial P}{\partial t} + \frac{\nabla_i \left( P\nabla^i {\cal S}\right)}{m} = 0~,
\end{equation}
respectively.


\section{The simplest solutions}

Let us consider the system (\ref{HJE}) and (\ref{cont}) in the simplest case when a typical particle of the 'universal' ensemble is at rest with respect to the universe and is considered to be free in the sense that $\nabla_i {\cal S} = V =0$. In this case the Hamilton's principal function can be expressed using the Machian energy (\ref{E}):
\begin{equation} \label{S0}
{\cal S}_0 = - Et + const~,
\end{equation}
and the system (\ref{HJE}) and (\ref{cont}) leads to the trivial continuity and Hamilton-Jacobi equations:
\begin{equation} \label{S_0}
\frac{\partial P_0}{\partial t} = 0~, ~~~~~ \frac{\partial {\cal S}_0}{\partial t} = - E ~.
\end{equation}
As the function $P_0$ is time-independent, it means that all the other $(N-1)$ particles of the ensemble are at rest with respect to the preferred frame.

In the case when only one particle moves with respect to the preferred frame, the Machian energy of the particle (\ref{E}) changes \cite{Gog2}, and the Hamilton's principal function (\ref{S0}), according to the definition (\ref{p}), takes the form:
\begin{equation} \label{S}
{\cal S} = p_ix^i - Et + const ~.
\end{equation}

Since all other $(N-1)$ particles are at rest, we again expect to have a stationary distribution function. Hence, the continuity equation (\ref{cont}) reduces to:
\begin{equation} \label{u+P}
\frac{\partial P}{\partial t} = - \frac 1m \left( \nabla^i {\cal S}
\nabla_i P + P \Delta {\cal S} \right) = 0 ~.
\end{equation}
In the case of a constant momentum
\begin{equation}
\Delta {\cal S} = \nabla_i p^i = 0~,
\end{equation}
and from (\ref{u+P}) it follows that there exists a constant co-momentum field orthogonal to $p^i$:
\begin{equation}
c^i \sim \frac{\nabla^i P}{P} ~, ~~~~~c^i p_i = 0~.
\end{equation}
Then the solution of the continuity equation (\ref{u+P}) can be written in the form:
\begin{equation} \label{P=P0}
P = P_0 e^{2 \oint_{L} dl_ic^i /\hbar}~,
\end{equation}
where $P_0$ is the distribution density in the case when all particles are at rest.

The situation reminds us the thermodynamical approach to quantum mechanics \cite{Gro}. The moving particle feels a resistance of the ensemble in the form of the heat flow $\Delta Q$. The distribution density for this non-equilibrium steady state can be written as:
\begin{equation}
P = P_0 e^{\Delta Q /kT} ~,
\end{equation}
where $k$ and $T$ are the Boltzmann constant and the absolute temperature of the ensemble, respectively. By equating the kinetic energy of the thermostat per degree of freedom, $kT/2$, with the average kinetic energy of an oscillator from the ensemble, $\hbar \omega/2$, and using the Boltzmann relation, $\Delta Q = 2\omega {\cal S}$, together with the description of the momentum (\ref{p}), we obtain the formula (\ref{P=P0}). This argument serves as another justification of the appearance of the factor $2/\hbar$ as the normalization constant in (\ref{P}).


\section{The quantum potential}

The exponential factor in (\ref{P=P0}) represents the perturbation of the density function $P_0$ due to the collective non-local response of the 'universal' ensemble to the motion of one particle. Alternatively, one can still use the unperturbed function $P_0$, instead of $P(t,x^i)$, but change the Hamilton principle function (\ref{S}) as follows:
\begin{equation} \label{S'}
{\cal S}' = \oint_{L} dl_ic^i - E't + const ~.
\end{equation}
This leads to the relation
\begin{equation} \label{S'P}
\nabla^i {\cal S}' = \frac \hbar 2 \frac{\nabla^i P}{P}
\end{equation}
and automatically 'gauges out' the second term in the continuity equation (\ref{cont}). In terms of the new function, (\ref{S'}), using the relation $\nabla_i c^i = 0$, the Hamilton-Jacobi equation (\ref{HJE}) takes the form:
\begin{equation} \label{HJE'}
\frac{\partial {\cal S}'}{\partial t} - V_q + V = 0 ~,
\end{equation}
where the second term coincides with the so-called quantum potential
\begin{equation} \label{Vq}
V_q = -\frac{\hbar^2}{2m} \frac{\Delta P}{P} ~,
\end{equation}
which appears in the models aiming at a classical interpretation of quantum mechanics \cite{Bohm}. This potential usually is interpreted as the source of uncertainty and non-locality that distinguish quantum systems from classical ones described by pure Hamilton-Jacobi equation. In our approach, because of the relation (\ref{S'P}), the change in the density function of one particle affects $V_q$ and hence, the motion of 'gravitationally entangled' particles of a Machian ensemble at any distance.

Note that the quantum potential (\ref{Vq}) is also useful to describe the so-called Fisher information of the probability density \cite{Fish}:
\begin{equation} \label{F}
I_F = \int d^3x \frac{\nabla_i P\nabla^i P}{P}~.
\end{equation}
Its connections with the Hamilton's principle was already noticed in \cite{Hamilton}.


\section{The Relativity Principle and complexity}

It is important to note that, since the universe in our model is considered to be finite, the system (\ref{HJE}) and (\ref{cont}) is not Galilean invariant. Namely, there exists a preferred frame where the average momentum of the world ensemble is zero,
\begin{equation} \label{}
\langle \nabla^i {\cal S} \rangle = \langle p^i \rangle = 0~.
\end{equation}
Hence, our description of the Machian universe actually uses the Aristotelian notion of space, where the distinguished state of motion is the rest. The Aristotelian space is also the underlying geometry of quantum theory \cite{Val}.

Let us explain now how one can introduce the relativity principle for the world ensemble. It is important to note that for the inertial particles the value of $V_q$ and $I_F$, and the form of the Hamilton-Jacobi equation (\ref{HJE'}), are independent of velocity. One can try to remove the term with the velocity from the expression of the Hamilton function (\ref{S}) (the first term in the right hand side) by introducing the Galilean symmetry
\begin{equation}\label{Gal}
t' = t~, ~~~~~x_i' = x_i - v_it~.
\end{equation}
Then the Hamilton function will be identical to the case of the particle at rest with respect to the universe. However, the distribution functions of the whole ensemble $P_v$ in this case changes its form due to those transformations and obeys the continuity equation with the non-vanishing flow:
\begin{equation} \label{cont-v}
\frac{\partial P_v}{\partial t} + \nabla_i \left( v^i P_v \right) = 0~.
\end{equation}
For the inertial particle we again (as in (\ref{u+P})) can restrict ourselves to the stationary state:
\begin{equation}
\frac{\partial P_v}{\partial t} = \nabla_i v^i = 0~.
\end{equation}
The difference from the case of a single moving particle is conceptual, because now it is assumed that the particle is at rest, and all other particles in the universe move with constant velocity $v^i \sim \nabla^i S_v$. Since the universe in our model is considered to be finite, the constant flow will result in the leakage of particles and hence, the information about the system, outside the horizon. This requires a modification of the variational principle by introducing the boundary terms. Then from (\ref{Ai}) it follows:
\begin{equation} \label{Sommer}
\oint_L dl^i \nabla_i {\cal S}_v = 2\pi n \hbar ~,
\end{equation}
where $n$ is the number of particles which cross any closed boundary $L$. The relation (\ref{Sommer}) coincides with the Bohr-Sommerfeld quantization condition in old quantum theory. Since $n$ is an integer number, this property can be formalized by introducing a single-valued complex function
\begin{equation} \label{psi}
\psi = e^{i S_v/\hbar} = e^{i \left( 2\pi n + S_v/\hbar \right)}~,
\end{equation}
so that (\ref{Sommer}) is automatically fulfilled.

Thus, the introduction of the relativity principle for the finite 'universal' ensemble described by the equations (\ref{HJE}) and (\ref{cont}) is equivalent to the introduction of the Bohr-Sommerfeld like quantization condition (\ref{Sommer}). It explains why the use of complex wave function is convenient and effective in the formalism of quantum mechanics.


\section{The Schr\"odinger equation}

It is known that a justification of the quantization condition (\ref{Sommer}) is problematic in the models which try to derive the Schr\"odinger equation using the concepts from classical physics \cite{Gro,Mad,Nel,Bohm,Ha-Re}. The classical derivations do not reproduce the quantum Schr\"odinger equation exactly unless this condition is imposed \cite{Wa-Ta}.

Our approach provides us with all necessary ingredients to derive the Schr\"odinger equation from our classical Machian system (\ref{HJE}) and (\ref{cont}). Using the Madelung transformation \cite{Mad},
\begin{equation} \label{Mad}
\Psi = \sqrt{P} \psi = e^{(K + i S)/\hbar}~,
\end{equation}
and the Bohr-Sommerfeld like quantization condition (\ref{Sommer}), one can reduce the system of non-linear equations (\ref{HJE}), (\ref{cont}) and (\ref{HJE'}) to the linear Schr\"odinger equation,
\begin{equation} \
i \hbar \frac{\partial \Psi}{\partial t} = \left(-\frac {\hbar^2}{2m}\Delta + V \right)\Psi~,
\end{equation}
for the complex wave function $\Psi$.

Thus, we have demonstrated that the equation of motion of a particle in our Machian model is identical to the Schr\"odinger equation which is postulated in the standard quantum theory.


\section{Conclusion}

In this paper the quantum behavior of matter in the Machian model of the universe is considered. The assumption that the collective long-range gravitational interaction of all particles in the universe gives rise to a non-local Machian potential allows us to consider all the particles in the universe as the members of a world statistical ensemble. The Planck constant is interpreted as a portion of the total action of the finite universe per one particle of the Machian ensemble. The Schr\"odinger equation is derived from the non-linear system of continuity and Hamilton-Jacobi equations which describe the movement of a particle of the Machian ensemble. The crucial points of the derivation are (i) the appearance of the quantization condition due to the discreetness of the ensemble and (ii) the introduction of the relativity principle in the Machian model with the preferred frame. The complex wave function is constructed from the pair of real classical functions: the probability density and the Hamilton's principal function. The model also suggests a physical interpretation of the components of the complex wave function and improves the traditional formulation of quantum mechanics where the wave functions themselves have no direct physical interpretation. We also note that the appearance of quantum probabilities in the model could be connected with the leaking of the information about the interacting particles outside the horizon.

\medskip


\noindent {\bf Acknowledgments:} This work was supported by the grant of Rustaveli National Science Foundation $ST~09.798.4-100$.


\end{document}